\documentclass[%
reprint,
showkeys,
showpacs,
nofootinbib,
amsmath,amssymb,
10pt]{revtex4-1}
\usepackage{graphicx}
\usepackage{dcolumn}
\usepackage{bm}
\usepackage[colorlinks,citecolor=blue,urlcolor=blue,linkcolor=blue]{hyperref}
\usepackage[figtopcap]{subfigure}
\begin{document}

\title{Exact Teleparallel Gravity of Binary Black Holes}
\author{W. El Hanafy$^{1,3}$}
\email{waleed.elhanafy@bue.edu.eg}
\author{G. G. L. Nashed$^{1,2,3}$}%
\email{nashed@bue.edu.eg}
\affiliation{$^{1}$Centre for theoretical physics, the British University in Egypt, 11837 - P.O. Box 43, Egypt}
\affiliation{$^{2}$Mathematics Department, Faculty of Science, Ain Shams University, Cairo, Egypt}
\affiliation{$^{3}$Egyptian Relativity Group (ERG)}
\begin{abstract}
An exact solution of two singularities in the teleparallel equivalent to general relativity  theory has been obtained. A holographic visualization of the binary black holes (BBHs) space-time, due to the non vanishing torsion scalar field, has been given. The acceleration tensor of BBHs space-time has been calculated. The results identify the repulsive gravity zones of the BBHs field. The total conserved quantities of the BBHs has been evaluated. Possible gravitational radiation emission by the system has been calculated without assuming a weak field initial data.
\end{abstract}

\maketitle
\section{Introduction}\label{S1}
The discovery of the first pulsars in a binary system PSR B1913+16 \cite{HT75} represents a natural laboratory for testing theories of gravity. As a matter of fact the parameterized post Newtonian technique recommends the GR theory to pass the binary pulsar laboratory test. However, the general relativistic exact solution of such systems is not recognized yet. The first trial was by \cite{Czn24} as an axially symmetric vacuum solution of two monopoles, the solution has been reobtained in \cite{S36}. A more extensive studies on the two body problem in GR has been studied \cite{C74,C75,CL77,CH79,C82,CL86}. But these solutions, in fact, are in a series trouble with GR theory. The so-called Curzon solution has been obtained as a vacuum solution of two singularities, while the solution violates the elementary flatness condition between the two monopoles. The only physical interpretation is that the region between the singularities cannot represent a vacuum solution of Einstein field equations by introducing an inert strut (i.e conical singularity) to keep the static configuration of the system balanced \cite{K1968,S85}, while other trials to remove the strut was by considering the spin-spin interaction \cite{DH82}, later it has been shown that double Kerr space-time cannot afford a balanced system \cite{GN87}. Also, the solution shows a peculiar singularity behaviour, whereas it exhibits a directional singularity allows the particles approaching the singularity along the axis of symmetry to be geodesically complete accessing to some new region of the space-time \cite{SM73}. This leads some to conclude that the singularities in Curzon solution are actually rings rather point-like particles \cite{F91,AFM91,AL12}. Interestingly, for the superposition of a ring and another body, a membrane-like singularity appears besides the strut one \cite{LO88,LO98}. Also, it has been shown that the interior of cylindrically symmetric space-times topology is physically acceptable, if there exists a negative energy cosmic string density \cite{L85}. Moreover, this solution has been modified to describe the pulsation time delay (Shapiro delay) of signals from binary pulsar systems by using a co-rotating coordinates \cite{WAH2010}.

In our study, we investigate possible solution of two monopoles in the TEGR framework. We expect new insights and interpretations when use different tools, i.e teleparallel torsion gravity, of such systems. One of the important features of the TEGR theory, that is the vacuum solutions do not imply the vanishing of the teleparallel torsion scalar, unlike the GR case in which the vanishing of the Ricci scalar in vacuum solutions is necessarily. So we expect the torsion scalar to play an important role to visualize the space-time, the more important role is to provide an alternative source of the compressional medium between the two singularities. We organize the article as follows: In Section \ref{S2}, we briefly review the Lagrangian formalism of TEGR theory. In Section \ref{S3}, we derive an exact solution of the TEGR field equations for a BBHs space-time. In Section \ref{S4}, we use the non vanishing torsion scalar to visualize the vacuum solution of the BBHs gravitational field. In Section \ref{S5}, we evaluate the acceleration tensor to determine the repulsive gravity zones of the BBHs field. In Section \ref{S6}, we calculate the conservative quantities of the BBHs space-time and an estimate value of its gravitational radiation emission. In Section \ref{S7}, we summarize and conclude the work.
\section{Teleparallel Equivalent of General Relativity}\label{S2}
The TEGR is an alternative description of gravitation of Einstein's GR. The theory has been used to reexamine many GR solutions \cite{N2006,N2007,N2010}. It is constructed from tetrad fields\footnote{The Greek indices describe the components of tangent
space to the manifold (space-time), while the Latin ones
describe the components of the space-time} $(e^{\mu}{_{i}})$ instead of a metric tensor fields $g_{\mu \nu}$. The metric tensor can be constructed from the tetrad fields: $g_{i j}=O_{\mu \nu}e^{\mu}{_i}e^{\nu}{_j}$, where $O_{\mu \nu}=\textmd{diag}(1,-1,-1,-1)$ is the Minkowski metric for the tangent space, so the Levi-Civita (symmetric) connection $\overcirc{\Gamma}{^{i}}{_{j k}}$ can be constructed. However, it is possible to construct Weitzenb\"{o}ck nonsymmetric connection $\Gamma^{i}{_{j k}}=e_{\mu}{^{i}}\partial_{j}e^{\mu}{_{k}}$ \cite{Wr}. The Weitzeinb\"{o}ck 4-space is described as a pair $(M,e_{\mu})$, where $M$ is an $4$-dimensional smooth manifold and $e_{\mu}$ ($\mu=0,\cdots, 3$) are $4$-linearly independent vector fields defined globally on $M$. The Weitzenb\"{o}ck space is characterized by the vanishing of the tetrad's covariant derivative, i.e. $\nabla_{_{j}}e^{\mu}{_{i}}\equiv 0$, where the covariant derivative $\nabla_{j}$ is with respect to (w.r.t.) the Weitzenb\"{o}ck connection. So this property identifies auto parallelism or absolute parallelism condition. As a matter of fact, the $\nabla_{j}$ operator is not covariant under $SO(1,3)$ group, i.e local  Lorentz transformations (LLT). The lack of the local Lorentz symmetry allows all LLT invariant geometrical quantities to rotate freely in every point of the space \cite{M2013}. In this sense, the symmetric metric (10 degrees of freedom) cannot predict exactly one set of tetrad fields; then the extra six degrees of freedom of the 16 tetrad fields need to be fixed in order to identify exactly one physical frame. It can be shown that the metricity condition is fulfilled as a consequence of the absolute parallelism condition. The Weitzenb\"{o}ck is curvature free while it has a non-vanishing torsion given by
\begin{equation}\label{torsion}
T^{i}{_{j k}} :=\Gamma^{i}{_{j k}}-\Gamma^{i}{_{k j}}.
\end{equation}
and contortion is given by
\begin{equation}\label{contortion}
{K^{i j}}_k :=-\frac{1}{2}\left({T^{i j}}_k-{T^{j i}}_k-{T_k}^{i j}\right),
\end{equation}
In the teleparallel space one may define three Weitzenb\"{o}ck invariants: $I_{1}=T^{i j k}T_{i j k}$, $I_{2}=T^{i j k}T_{j i k}$ and $I_{3}=T^{i}T_{i}$, where $T^{i}=T_{j}{^{i j}}$. We next define the invariant $T=AI_{1}+BI_{2}+CI_{3}$, where $A$, $B$ and $C$ are arbitrary constants \cite{M2013}. For the values: $A=1/4$, $B=1/2$ and $C=-1$  the invariant $T$ is just the Ricci scalar $R^{(\overcirc{\Gamma})}$, up to a total derivative term as we will show below; then a teleparallel version of gravity equivalent to GR can be achieved. The teleparallel torsion scalar is given in the compact form
\begin{equation}\label{Tor_sc}
T := {T^i}_{j k}{S_i}^{j k},
\end{equation}
where the superpotential tensor ${S_i}^{j k}$ is defined as
\begin{equation}\label{Stensor}
{S_i}^{j k}:= \frac{1}{2}\left({K^{j k}}_i+\delta^j_i{T^{l k}}_l-\delta^k_i{T^{l j}}_l\right),
\end{equation}
which is skew symmetric in the last two indices. We next highlight some useful relations between Riemannian and teleparallel geometries. The double contraction of the first Bianchi identity of the teleparallel geometry gives
\begin{equation}\label{Bianchi1}
R^{(\overcirc{\Gamma})}=-T^{(\Gamma)}-2\overcirc{\nabla}_{i}T^{j i}{_{j}},
\end{equation}
where the covariant derivative $\overcirc{\nabla}$ is w.r.t. the Levi-Civita connection $\overcirc{\Gamma}$. The second term in the right hand side is a total derivative. So, it has no contribution in the variation when use the right hand side instead of the Ricci scalar in the Einstein-Hilbert action. Consequently the Ricci and teleparallel torsion scalars are equivalent up to a total derivative term. In spite of this quantitative equivalence they are qualitatively different. For example, the Ricci scalar is invariant under LLT while the total derivative term is not, so the torsion scalar. Accordingly, the TEGR \textit{Lagrangian} is not invariant under LLT \cite{1010.1041,1012.4039,KS2015}.

The TEGR gives the action a gauge gravitational field Lagrangian as \cite{M94,M2002}
\begin{equation}\label{action}
\mathcal{S}=\frac{M_{\textmd{\tiny Pl}}^2}{2}\int |e| \left[T+ \mathcal{L}_{M}(\Phi_{A})\right]~d^{4}x,
\end{equation}
where $\mathcal{L}_{M}$ is the Lagrangian of the matter fields $\Phi_{A}$ and $M_{\textmd{\tiny Pl}}$ is the reduced Planck mass, which is related to the gravitational constant $G$ by $M_{\textmd{\tiny Pl}}=\sqrt{\hbar c/8\pi G}$. Assume the units in which $G = c = \hbar = 1$. In the above equation, $|e|=\sqrt{-g}=\det\left({e^\mu}_i\right)$. The variation of (\ref{action}) w.r.t. the tetrad fields ${e^\mu}_i$ give rise to the following field equations \cite{M94}
\begin{equation}\label{TEGR}
\partial_{j}(e S_{\mu }{^{i j}})=4\pi e e_{\mu}{^j}(t_{j}{^{i}}+\Theta_{j}{^{i}}),
\end{equation}
where $S_{\mu}{^{i j}}=e_{\mu}{^l}S_{l}{^{i j}}$, the (pseudo) tensor $t_{i}{^j}$ is
\begin{equation}\label{grav_EM_tensor}
t_{i}{^{j}}=\frac{1}{16\pi}(4T^{l}{_{m i}}S_{l}{^{j m }}-\delta^{j}_{i}T),
\end{equation}
and the matter energy-momentum tensor is
\begin{equation}\label{EM_tensor}
\Theta_{i}{^j}=e^{\mu}{_{i}}\left(-\frac{1}{e}\frac{\delta L_{m}}{\delta e^{\mu}{_{j}}}\right).
\end{equation}
As the tensor $S_{\mu}{^{i j}}$ is skew-symmetric, i.e $S_{\mu }{^{i j}}=-S_{\mu}{^{j i}}$, this implies that
$\partial_{i}\partial_{j}(e S_{\mu}{^{i j}})\equiv 0$ \cite{M2013}. Therefore,
$$\partial_{j}\left[e e_{\mu}{^i}(t_{i}{^{j}}+\Theta_{i}{^{j}})\right]=0.$$
The pseudo tensor $t_{i}^{j}$ has no equivalent in GR theory. To reveal its nature we find that the above equation gives rise to the continuity equation
$$\frac{d}{dt}\int_{V}e e_{\mu}{^i}(t_{0\, i}+\Theta_{0\, i})d^3 x=-\oint_{\Sigma}\left[e e_{\mu}{^i}(t_{\nu i}+\Theta_{\nu i})\right]d\Sigma^{\nu},$$
where the integration is on a three dimensional volume $V$ bounded by the surface $\Sigma$. This leads to interpret $t_{i}{^j}$ as the energy-momentum tensor of the gravitational field \cite{M94,BS2015}.
\section{Exact Teleparallel Two Body Problem}\label{S3}
The metric characterizing the space, with axial-symmetric static gravitational field in cylindrical coordinates ($t,z,\rho,\varphi$) about the $z$ axis can be written as
\begin{equation}\label{axial_metric}
ds^{2}=e^{\nu}dt^{2}-e^{\mu}\left( dz^{2}+d\rho^{2}\right) -e^{-\nu}\rho^{2}d\varphi^{2},
\end{equation}
where $\mu\equiv \mu\left(z,\rho\right),~ \nu \equiv \nu\left(z,\rho\right)$ and the coordinate ranges are
$$\infty>t>-\infty,\quad \infty>z>-\infty,\quad \infty>\rho\geq 0,\quad 2\pi\geq \varphi\geq 0.$$
The metric (\ref{axial_metric}) gives rise to the following diagonal tetrad
\begin{equation}\label{tetrad}
e^{\mu}{_{i}}=\textmd{diag}(e^{\nu/2},~e^{\mu/2},~e^{\mu/2},~\rho e^{-\nu/2}).
\end{equation}
The field equations (\ref{TEGR}) read non-vanishing components of the energy-momentum tensor as
\begin{equation}\label{T00}
    \Theta_{0}{^{0}}=\frac{e^{-\mu}}{32\pi}\left[\nu_{z}^2+\nu_{\rho}^2+2(\mu_{z z}-\nu_{z z}+\mu_{\rho\rho}-\nu_{\rho\rho})-\frac{4\nu_{\rho}}{\rho}\right].
\end{equation}
\begin{equation}\label{T11}
    \Theta_{1}{^{1}}=-\Theta_{2}{^{2}}=\frac{e^{-\mu}}{32\pi\rho}\left(2\mu_{\rho}+2\nu_{\rho}+\rho\nu_{z}^{2}-\rho\nu_{\rho}^{2}\right),
\end{equation}
\begin{equation}\label{T12}
    \Theta_{1}{^{2}}=\Theta_{2}{^{1}}=\frac{e^{-\mu}}{16\pi\rho}\left(\mu_{z}+\nu_{z}-\rho\nu_{z}\nu_{\rho}\right),
\end{equation}
\begin{equation}\label{T33}
    \Theta_{3}{^{3}}=-\frac{e^{-\mu}}{32\pi}\left(\nu_{z}^2+2\nu_{z z}+2\mu_{z z}+\nu_{\rho}^2+2\mu_{\rho\rho}+2\nu_{\rho\rho}\right),
\end{equation}
where $\mu_{z}=\frac{\partial \mu}{\partial z}$, $\mu_{\rho}=\frac{\partial \mu}{\partial \rho}$, $\mu_{zz}=\frac{\partial^{2} \mu}{\partial z^{2}}$; and so on. We solve the system for vacuum $\Theta_{i}{^{ j}}=0$. We combine
\begin{equation}\label{laplace}
    \Theta_{0}{^{0}}-\Theta_{3}{^{3}}=\frac{e^{-\mu}}{8\pi}\left(\nu_{z z}+\nu_{\rho\rho}+\nu_{\rho}/\rho\right)=0.
\end{equation}
The vanishing of the quantity between the brackets in (\ref{laplace}) is just the Laplace equation $\nabla^{2}\nu=0$ in cylindrical coordinates with an axis of symmetry about $z$. The field equation (\ref{laplace}) gives the linear Newtonian gravity; then the field of two monopoles can be introduced as a superposition of the gravitational field of two singularities as
\begin{equation}\label{nu}
    \nu(z,\rho)=-\frac{2m_{1}}{r_{-}}-\frac{2m_{2}}{r_{+}},
\end{equation}
where $m_{1}$ and $m_{2}$ are two constants of integration, $r_{-}$ and $r_{+}$ are bipolar coordinates of an observer in ($z,\rho$) plane referred to two singularities, respectively, separated by a distance of $2d$ with the origin of a reference frame lies at the midpoint
between the two singularities on the $z$-axis. The direction of the two singularities can be defined using the bipolar coordinates as
\begin{equation}\label{bipolar}
    r_{\pm}^{2} =\left(z \pm d \right)^{2}+\rho^{2}.
\end{equation}
The linearization of the GR tells that $g_{00}$ is the Newtonian potential; then the constants $m_{1}$ and $m_{2}$ represent masses. We introduce the function $\lambda=\nu+\mu$, such that the field equation (\ref{T12}) reads
\begin{equation}\label{mu_z}
    \Theta_{1}{^{2}}=0 \Rightarrow \lambda_{z}=\rho\nu_{\rho}\nu_{z},
\end{equation}
while (\ref{T11}) gives rise to
\begin{equation}\label{mu_rho}
    \Theta_{1}{^{1}}=0 \Rightarrow \lambda_{\rho}=-\frac{1}{2}\rho\nu_{z}^2+\frac{1}{2}\rho\nu_{\rho}^2.
\end{equation}
Combining (\ref{mu_z}) and (\ref{mu_rho}), and substituting from (\ref{nu}) we evaluate
\begin{eqnarray}\label{lambda}
\nonumber    \lambda(z,\rho)&=&-\left[\frac{m_{1}^{2}}{r_{-}^{4}}+\frac{m_{2}^{2}}{r_{+}^{4}}\right]
\rho^{2}+\frac{m_{1}m_{2}}{d^{2}}\left[\left({z^{2}+\rho^{2}-d^{2}}\over{r_{-}~r_{+}}\right)-1\right],
\end{eqnarray}
where the constants of integration have identified by imposing the boundary condition: the space is asymptotically flat. Consequently, we have
\begin{eqnarray}\label{mu}
\nonumber    \mu(z,\rho)&=&2\left[\frac{m_{1}}{r_{-}}+\frac{m_{2}}{r_{+}}\right]
-\left[\frac{m_{1}^{2}}{r_{-}^{4}}+\frac{m_{2}^{2}}{r_{+}^{4}}\right]
\rho^{2}\\
&+&\frac{m_{1}m_{2}}{d^{2}}\left[\left({z^{2}+\rho^{2}-d^{2}}\over{r_{-}~r_{+}}\right)-1\right].
\end{eqnarray}
As mentioned above the contribution of the classical theory of gravity is represented by the linear equation (\ref{nu}), while the non-linear contribution is given by the quadratic terms in equation (\ref{mu}). The obtained solution is equivalent to the general relativistic one \cite{Czn24}. It has been pointed out that the space (\ref{axial_metric}) violates the elementary flatness in the region between the two masses \cite{ER36,S85}. This can be seen when taking a circle with a center on the $z$-axis in the subspace $z=constant$, $t=constant$ with $|z|<d$, one finds that the ratio of the circumference $C$ to the radius $R$ of that circle as $R\rightarrow 0$ is $C/R \rightarrow 2\pi e^{\lambda(z,0)}$. This requires that $\lambda(z,0)$ to vanish everywhere on the plane which is not fulfilled for the $z$-axis region with $|z|<d$, accordingly the ratio $C/R$ does not approach $2\pi$. This type of singularity is called conical singularity, the only physical interpretation from the GR point of view is to assume some sort of matter with $T_{\mu\nu} \neq 0$ within this region, i.e. to introduce a \textit{strut} with zero active gravitational mass carrying the stress between the two masses \cite{K1968,S85}. However, the TEGR solution might enable to examine the role of the teleparallel gravity to reveal the nature of the repulsion between the two singularities.
\section{Visualization of BBHs space-time}\label{S4}
In this section we investigate the singularities and horizons of the above solution via curvature verses torsion invariants. We evaluate the Ricci scalar of the BBHs space-time described by the induced metric (\ref{axial_metric})
\begin{equation}\label{Rsc}
R(z,\rho)=-\frac{e^{-\mu}}{2 \rho}(2\rho\mu_{\rho\rho}+2\rho\mu_{zz}+\rho\nu_{z}^2-2\nu_{\rho}+\rho\nu_{\rho}^2).
\end{equation}
It is clear that the solution given by (\ref{nu}) and (\ref{mu}) implies the vanishing of the Ricci scalar as expected for a vacuum solution. Recalling that the energy-momentum component (\ref{T33}) of vacuum and by comparison with (\ref{Rsc}), we find that the Ricci scalar is enforced to produce the Laplace equation $\nabla^{2}\nu=0$ in vacuum. We conclude that the vanishing of Ricci scalar is natural for the vacuum choice. Consequently, all the invariants relevant to the Ricci would vanish and do not afford visual images of the space-time. Moreover, the solution is static, we expect all the magnetic part of Weyl tensor to vanish. This left us with only two possible non-vanishing invariants\footnote{It is worth mentioning that the so called Kretschmann scalar is just $8W1R$ in this space \cite{AL12}.}, that are
\begin{eqnarray}
\nonumber    W1R&=&\frac{e^{-2\mu}}{24\rho}\left[3\nu_{\rho\rho}\rho^2-6\rho\nu_{\rho\rho}(\mu_{\rho}\nu_{\rho}\rho+\nu_{zz}\rho
-\mu_{z}\nu_{z}\rho-\nu_{\rho}-\mu_{\rho})\right.\\
\nonumber &+&3\nu_{zz}^{2}\rho^2-6\rho\nu_{zz}\left\{(1-\rho\mu_{\rho})\nu_{\rho}+\mu_{z}\nu_{z}\rho+\mu_{\rho}\right\}+\mu_{\rho\rho}^{2}\rho^2\\
\nonumber
&+&2\rho\mu_{\rho\rho}(-\nu_{\rho}^{2}\rho+\mu_{z z}\rho-\nu_{z}^{2}\rho+2\nu_{\rho})+12\rho^{2}\nu_{\rho z}^{2}+\mu_{zz}^{2}\rho^{2}\\
\nonumber
&-&12\rho\nu_{\rho z}(-\nu_{z}-\mu_{z}+\rho\nu_{z}\mu_{\rho}+\mu_{z}\nu_{\rho}\rho)+\nu_{\rho}^{4}\rho^{2}-4\nu_{\rho}^{3}\rho\\
\nonumber
&-&2\rho\mu_{zz}(\nu_{\rho}^{2}\rho-2\nu_{\rho}+\nu{z}^{2}\rho)+\nu_{z}^4 \rho^2+3\mu_{z}^2+3\mu_{\rho}^2+6\mu_{z}\nu_{z}\\
\nonumber
&+&\nu_{\rho}^2(2\nu_{z}^2 \rho^2+7-6\mu_{\rho}\rho+3\mu_{\rho}^2\rho^2+3\mu_{z}^2 \rho^2)+\nu_{\rho}\left(-6 \mu_{\rho}^2 \rho\right.\\
\nonumber
&+&\left.\left.6 \mu_{\rho}-4 \nu_{z}^2 \rho-6 \mu_{z}^2 \rho\right)+\nu_{z}^{2}(3-6\mu_{\rho}\rho+3\mu_{\rho}^2 \rho^2+3\mu_{z}^2 \rho^2)\right],\\
&&\label{w1r}
\end{eqnarray}
and
\begin{eqnarray}
\nonumber    W2R&=&\frac{-e^{-3\mu}}{288\rho^3}\left[
\mu_{\rho\rho}^2\rho^2+2\rho(-\nu_{\rho}^2\rho+\mu_{zz}\rho-\nu_{z}^2\rho+2\nu_{\rho})\mu_{\rho\rho}\right.\\
\nonumber
&-&9\nu_{\rho\rho}^2\rho^2
+18\rho(\mu_{\rho}\nu_{\rho}\rho+\nu_{zz}\rho-\mu_{z}\nu_{z}\rho-\nu_{\rho}-\mu_{\rho})\nu_{\rho\rho}\\
\nonumber
&+&\mu_{zz}^2\rho^2
-2\rho(\nu_{\rho}^2\rho-2\nu_{\rho}+\nu_{z}^2\rho)\mu_{zz}-9\nu_{zz}^2\rho^2\\
\nonumber
&+&(18\left\{(-\mu_{\rho}\rho+1)\nu_{\rho}
+\mu_{z}\nu_{z}\rho+\mu_{\rho}\right\})\rho\nu_{zz}-36\rho^2\nu_{\rho z}^2\\
\nonumber
&+&36\rho(-\nu_{z}-\mu_{z}+\rho\nu_{z}\mu_{\rho}+\mu_{z}\nu_{\rho}\rho)\nu_{\rho z}+\nu_{\rho}^4\rho^2-4\nu_{\rho}^3\rho\\
\nonumber
&+&(-5-9\mu_{z}^2\rho^2+2\nu_{z}^2\rho^2-9\mu_{\rho}^2\rho^2
+18\mu_{\rho}\rho)\nu_{\rho}^2-9\mu_{z}^2\\
\nonumber
&+&(18\mu_{z}^2\rho-18\mu_{\rho}-4\nu_{z}^2\rho+18\mu_{\rho}^2\rho)\nu_{\rho}
+\nu_{z}^4\rho^2-9\mu_{\rho}^2\\
\nonumber
&+&\left.(-9+18\mu_{\rho}\rho-9\mu_{z}^2\rho^2-9\mu_{\rho}^2\rho^2)\nu_{z}^2-18\mu_{z}\nu_{z}
\right]\times\\
&&(\mu_{\rho\rho}\rho+\mu_{z z}\rho+2\nu_{\rho}-\nu_{\rho}^2\rho-\nu_{z}^2\rho).\label{w2r}
\end{eqnarray}
Substituting from (\ref{nu}) and (\ref{mu}) into (\ref{w1r}) and (\ref{w2r}), a straightforward calculations indicate that the the invariants $W1R$ and $W2R$ diverge when $r_{\pm}\rightarrow 0$. Since the solution has a curvature singularity at $r_{\pm} = 0$ but $g_{00}$ does not vanish for $r_{\pm} > 0$, the solution has no horizons. So the singularities at $r_{\pm}=0$ are physical and naked.

On the other hand, the three independent Weitzenb\"{o}ck invariants $I_{1}=T^{i j k}T_{i j k}$, $I_{2}=T^{i j k}T_{j i k}$ and $I_{3}=T^{i}T_{i}$, where $T^{i}=T_{j}{^{i j}}$ might help to analyze the singularities. We omit using these invariants separately in this study due to the lack of invariance under LLT. However, their combination gives a well defined invariant, that is the teleparallel torsion scalar \cite{CGSV2013}. The contracted Bianchi identity (\ref{Bianchi1}) shows that if we want the teleparallel torsion scalar to vanish this implies the vanishing of the total derivative term as well. As a matter of fact, this cannot be happen, at least, for this solution. We show this in more details, substituting from (\ref{tetrad}) into (\ref{torsion}) and (\ref{Stensor}), we evaluate the teleparallel torsion scalar
\begin{equation}\label{Tsc}
    T(z,\rho)=\frac{e^{-\mu}}{2\rho}\left(2\mu_{\rho}+2\nu_{\rho}-\rho\nu_{z}^{2}-\rho\nu_{\rho}^{2}\right).
\end{equation}
\begin{figure}[t]
\centering
\subfigure[~3D plot of the torsion scalar]{\label{fig1a}\includegraphics[scale=.5]{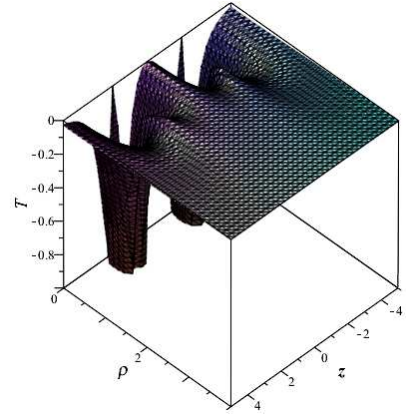}}\hspace{1cm}
\subfigure[~Torsion contour]{\label{fig1b}\includegraphics[scale=.5]{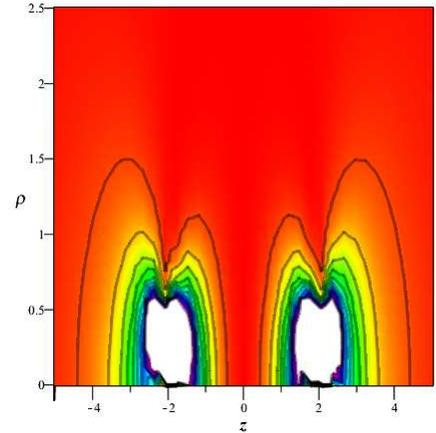}}
\caption[figtopcap]{Visualization of BBHs space-time: \subref{fig1a} a Holographic visualization of the teleparallel torsion scalar field formed by BBHs gravitational field;
\subref{fig1b} The contour lines of the teleparallel torsion show a repulsive zone in the inner region of the BBHs. The constants have been chosen as $m_1=1$, $m_2=1$ and $d=2$.}
\label{Fig1}
\end{figure}
Recalling that the energy-momentum component (\ref{T11}) of vacuum and by comparison with (\ref{Tsc}), the teleparallel torsion scalar of binary black holes reduces to
\begin{equation}\label{Tsc2}
    T(z,\rho)=-\nu_{z}^2e^{-\mu}\neq 0.
\end{equation}
The teleparallel torsion scalar vanishes if $\nu_{z}=0$. Consequently, the potential function $\nu$ becomes a function of the coordinate $\rho$ only, which contradicts the vacuum solution. So we conclude that the vacuum solution in fact prevents the torsion scalar to vanish completely unlike to the Ricci scalar.

Substituting from (\ref{nu}) and (\ref{mu}) into (\ref{Tsc2}), we find that the torsion scalar diverges when $r_{\pm}\rightarrow 0$ which is in agreement with the Weyl invariants $W1R$ and $W2R$ results. However, the study via the teleparallel torsion invariant $T$ is much simpler than $W1R$ and $W2R$ invariants. The non-vanishing torsion scalar of vacuum solutions, in general, might help to visualize the space-time properties. The case here of the BBHs vacuum solution can be seen in Figure \ref{Fig1}. Recognizably, the contours of the torsion scalar similar to the magnetic field lines of binary pulsars. However, the inner region shows clearly a repulsive behaviour of the torsion field lines. This recommends the non-vanishing teleparallel torsion scalar field to interpret the compression force that keeps the two singularities apart. Finally, we find that the singularity analysis of the BBHs via curvature verses torsion are in agreement as expected in the non charged or TEGR cases \cite{CGSV2013}. In the following section we use the acceleration tensor to investigate the repulsive gravity zones of the BBHs field.
\section{Repulsive Gravity Zones}\label{S5}
Consider the vierbein fields as reference frames of an observer in space-time defined by an arbitrary timelike worldling $C$. These frames can be characterized in an invariant way by antisymmetric acceleration tensor $\phi_{\mu \nu}$. This tensor generalizes the inertial accelerations of the frame, in analogy with Farady tensor, the acceleration tensor can identify $\phi_{\mu \nu}\rightarrow (\mathbf{a},\mathbf{\Omega})$, where $\bf a$ is the translational acceleration ($\phi_{(0)(\mu)}=a_{(\mu)}$) and $\bf \Omega$ is the frequency of rotation of the local spatial frame w.r.t. non-rotating frame \cite{M14}.

Let the worldline $C$ of the observer be denoted by $x^\mu(\tau)$, where $\tau$ is the proper time of the observer, whose  frame is adapted such that its components identify the velocity and acceleration along $C$ respectively by $e_{(0)}\,^i=u^i=dx^i/d\tau$ and $a^i=Du^i/d\tau$. The absolute derivative $D/d\tau$ is w.r.t. the Christoffel symbols $\overcirc{\Gamma}^i_{ j k}$ as
\begin{equation}
a^i={{Du^i}\over{d\tau}} = u^j \nabla_j e_{(0)}\,^i
= {{d^2 x^i}\over {d\tau^2}}+\,\,\overcirc{\Gamma}^i_{j k}
{{dx^j }\over{d\tau}} {{dx^k}\over{d\tau}}\,.
\label{4}
\end{equation}
We see that if $u^i=e_{(0)}\,^i$ represents a geodesic trajectory, then the frame is in free fall and $a^i=0=\phi_{(0)(\mu)}$. Therefore we conclude that non-vanishing values of the latter quantities represent inertial accelerations of the frame.
\begin{equation}
a^i= {{Du^i}\over{d\tau}} ={{De_{(0)}\,^i}\over {d\tau}} =
u^j \nabla_j e_{(0)}\,^i\,.
\label{accel-vect}
\end{equation}
Assuming that the observer carries an orthonormal tetrad frame $e_\mu\,^i$, the acceleration
of the frame along the path is given by
\begin{equation}
{{D e_\mu\,^i} \over {d\tau}}=\phi_\mu\,^\nu\,e_\nu\,^i\,,
\label{2}
\end{equation}
where $\phi_{\mu \nu}$ is the antisymmetric acceleration tensor. It follows from Eq. (\ref{2}) that
\begin{equation}
\phi_\mu\,^\nu= e^\nu\,_i {{D e_\mu\,^i} \over {d\tau}}=
e^\nu\,_i \,u^j\nabla_j e_\mu\,^i\,.
\label{3}
\end{equation}
The acceleration vector $a^i$ defined by Eq. (\ref{accel-vect}) may be projected on a frame yielding
$$a^\mu= u\;e^\mu\,_i\; \,a^i= \,e^\mu\,_i\, u^j \,\nabla_j e_{(0)}\,^i=\,\phi_{(0)}\,^\mu.$$
It is possible to rewrite the acceleration tensor in the form
\begin{equation}
\phi_{\mu \nu}={1\over 2} \lbrack T_{(0)\mu \nu}+T_{\mu (0)\nu}-T_{\nu(0)\mu}
\rbrack\,,
\label{5}
\end{equation}
where $T_{\mu \nu \rho}=e{_\nu}{^i} e{_\rho}{^j} T_{\mu ij}$, and $T_{\mu i j}=\partial_i e_{\mu j}-\partial_j e_{\mu i}$ is the
torsion tensor of the Weitzenb\"ock space-time. The expression of $\phi_{\mu \nu}$ is invariant under coordinate transformations \cite{M14}.
\begin{figure}[t]
\centering
\subfigure[~$T_{(0)(0)(1)}$ component]{\label{fig2a}\includegraphics[scale=.4]{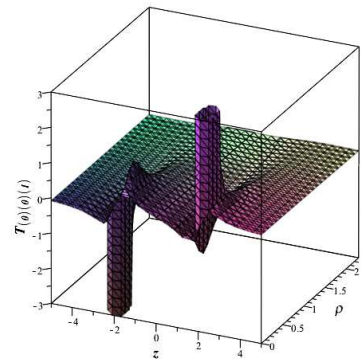}}
\subfigure[~$T_{(0)(0)(2)}$ component]{\label{fig2b}\includegraphics[scale=.4]{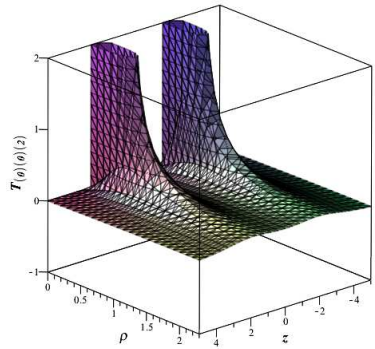}}
\caption{acceleration tensor components:
\subref{fig2a} along the $z$-direction have some regions with negative values near the singularities and between the two singularities;
\subref{fig2b} along the $\rho$-direction have only positive values. The constants have been chosen as $m_1=1$, $m_2=1$ and $d=2$.}
\label{Fig2}
\end{figure}
The values of the 6 components of the acceleration tensor may be used to characterize the frame, since $\phi_{\mu \nu}$ is not invariant under
local SO(3,1) (Lorentz) transformations. Alternatively, the frame may be characterized (i) by the identification $u^i=e_{(0)}\,^i$ (this equation fixes 3 components, because $e_{(0)}\,^0$ is fixed by normalization), and (ii) by the 3 orientations in the three-dimensional space of the components $e_{(1)}{^i},e_{(2)}{^i},e_{(3)}{^i}$.

For a frame that undergoes the usual translational and/or rotational accelerations in flat space-time, $\phi_{\mu \nu}$ yields the expected, ordinary values. An interesting application of the acceleration tensor is the following.

Let us consider a static observer in the BBHs space-time described in the cylindrical coordinates ($t,z,\rho,\varphi$) by the vierbein (\ref{tetrad}). The frame of the observer must satisfy $e_{(0)}\,^i=0=u^i$, the translation accelerations identify acceleration tensor components
$$\phi_{(0)(\mu)} = T_{(0)(0)(\mu)} = e_{(0)}{^{i}} e_{(\mu)}{^{j}} T_{(0)i j},$$
where the non-vanishing components of the translation acceleration tensor of the BBHs space-time are
$$\phi_{(0)(1)}=\frac{1}{2}\nu_{z}e^{-\mu/2},\quad \phi_{(0)(2)}=\frac{1}{2}\nu_{\rho}e^{\mu/2}.$$
The translation inertial acceleration plots, Figure \ref{Fig2}, indicate that there are  repulsive gravity regions along $z$-axis which is identified by $\phi_{(0)(1)}<0$ values, while they indicate that the gravitational acceleration is always attractive along $\rho$-axis. In this case the translational acceleration vector  ${\bf a}\equiv  \left(\phi_{(0)(1)},\phi_{(0)(2)},0,0\right)$ can be given by
$$\mathbf{a}=\frac{1}{2}\nu_{z}e^{-\mu/2} \hat{\mathbf{z}}+\frac{1}{2}\nu_{\rho}e^{-\mu/2}\hat{\rho},$$
where $\hat{\mathbf{z}}$ and $\hat{\rho}$ are unit vectors along $z$ and $\rho$ directions, respectively. Consequently,
\begin{eqnarray}
\nonumber a(z,\rho)&=&\frac{1}{2}e^{-\mu/2}\left(\nu_{z}^{2}+\nu_{\rho}^{2}\right)^{1/2}\\
&=&2\left(\frac{m_{1}^{2}}{r_{-}^{4}}+\frac{m_{2}^{2}}{r_{+}^{4}}\right)^{1/2}e^{-\mu/2}.
\end{eqnarray}
For simplicity we take $m_{1}\simeq m_{2}$, along the $z$-axis outside the BBHs, i.e. $|z|\gg d$. The acceleration is
\begin{eqnarray}
\nonumber a_{\textmd{out}}&=&-2\frac{(m_{1}+m_{2})z d}{(z+d)^2(z-d)^2}e^{-\mu(z,0)/2},\\[3pt]
\nonumber      &\sim&-2\frac{m_{1}m_{2}}{z^2}.
\end{eqnarray}
Also, its value in the inter-medium region, i.e. $|z|\ll d$, is given by
\begin{eqnarray}
\nonumber a_{\textmd{in}}&=&2\frac{(m_{1}+m_{2})z d}{(z+d)^2(z-d)^2}e^{-\mu(z,0)/2},\\[3pt]
\nonumber      &\sim&2\frac{m_{1}m_{2}}{z^2}.
\end{eqnarray}
\begin{figure}[t]
\begin{center}
\includegraphics[scale=0.5]{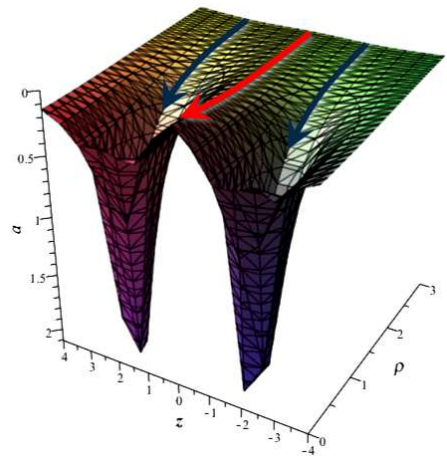}
\caption{The acceleration in the BBHs gravitational field shows a deceleration region between the two singularities. This indicates a repulsive gravity region. The constants have been chosen as $m_1=1$, $m_2=1$ and $d=2$.}\label{acc_bin}
\end{center}
\end{figure}
The above expressions represent the inertial acceleration needed to cancel exactly the gravitational acceleration maintaining  the frame static. It is clear that the outer regions of the BBHs along $z$-axis experience the usual attractive gravity, while the inner region indicates a repulsive gravity zone. We examine the acceleration along $\rho$-axis, we get
$$a(0,\rho)=\frac{(m_{1}+m_{2})\rho}{(\rho^2+d^2)^{3/2}}e^{-\mu(0,\rho)/2}\sim -\frac{m_{1}+m_{2}}{\rho^{2}}.$$
The last term is evaluated where $\rho \gg d$, which indicates that the gravity along the $\rho$-axis is always attractive. We also interested to examine the acceleration at the center of mass, which can be given as
$$a(0,0)=\frac{m_{1}-m_{2}}{d^2}e^{-\mu(0,0)/2}.$$
The inertial acceleration at the center vanishes if $m_{1}=m_{2}$. This is expected as the gravitational acceleration vanishes at the same condition so that the inertial acceleration must vanish to maintain the frame static. In Figure \ref{acc_bin}, we plot the acceleration in ($z$,$\rho$) plane of the BBHs. The plot shows that acceleration increases as an observer falls towards any of the two monopoles, while it decreases at some regions between them. This deceleration regions identify the repulsive gravity zones of the BBHs space-time.
\section{Conserved Quantities of the Binary Solution}\label{S6}
The energy localization problem has been studies within the tetrad formalism in many literature \cite{SNH1996,SN1997}. The Lagrangian of the TEGR using forms techniques takes the form\footnote{The role of the addition of non-
Riemannian parity odd pseudoscalar curvature to the Hilbert-Einstein-Cartan scalar curvature has been investigated   (cf., \cite{M09,M091,M092} and references therein).}
\begin{equation}\label{g1}
{\cal V}=-\frac{1}{16 \pi}{\cal T}^\alpha \wedge ^\ast \left({^ {\tiny{(
1)}}{\cal T}_\alpha}-2~{^{\tiny{( 2)}}{\cal T}_\alpha} -\frac{1}{2}~{^ {\tiny{(
3)}}{\cal T}_\alpha} \right),
\end{equation}
where $\ast$ denotes the Hodge duality in the metric $g_{\alpha \beta}$ which is assumed to be flat Minkowski metric $g_{\alpha \beta}=O_{\alpha\beta}=diag(+1,-1,-1,-1)$, that is used to raise and lower local frame (Greek) indices. The variation of the total action w.r.t. the coframe gives the equation of motions in the  from \cite{LOP}
\begin{equation}\label{g2}
D{\cal H}_\alpha-{\cal E}_\alpha=\Sigma_\alpha, \ \ \textmd{where} \ \ \ { \Sigma}_\alpha
\stackrel {\rm def.}{=} \frac{\delta  {\cal{ L}}_{mattter}}{\delta
\vartheta^\alpha},
\end{equation}
that is the canonical energy-momentum current 3-form of matter  which is the source. Respect to the general Lagrange-Noether scheme \cite{HMMN}, one can derive from (\ref{g1}) the translational momentum 2-form and the canonical energy-momentum 3-form:
\begin{eqnarray}
{\cal H}_{\alpha} &:=& -\frac{\partial {\cal V}}{\partial {\cal T}^\alpha}=\frac{1}{8\pi}
\ast \left({^  {\tiny{( 1)}} {\cal T}_\alpha}-2{^  {\tiny{( 2)}} {\cal T}_\alpha}
-\frac{1}{2}{^  {\tiny{( 3)}} {\cal T}_\alpha} \right), \\
 {\cal E}_\alpha
&:=& \frac{\partial {\cal V}}{\partial
\vartheta^\alpha}=e_\alpha \rfloor {\cal V}+\left(e_\alpha \rfloor
{\cal T}^\beta \right) \wedge {\cal H}_\beta.\label{g3} \end{eqnarray} Because of the geometric identities \cite{ORP}, the Lagrangian (\ref{g1}) can be rewritten in the form  \begin{equation}\label{g4}
 {\cal V}=-\frac{1}{2} {\cal T}^\alpha\wedge {\cal H}_\alpha.\end{equation} The existence  of the
connection field ${\Gamma^\alpha}_\beta$ plays an important role in the
regularizing due to the following: \vspace{.3cm}\\
\underline{i}) The TEGR theory be invariant under the LLT of the coframe, i.e.,
the Lagrangian (\ref{g1}) is covariant under the change of
variables
\begin{eqnarray}\label{g5}
& &  \vartheta'^\alpha={\Lambda^\alpha}_\beta \vartheta^\beta, \nonumber\\
&&
{ \Gamma'_\alpha}^\beta={\Lambda^\mu}_\alpha {\Gamma_\mu}^\nu {(\Lambda^{-1})^\beta}_\nu-{(\Lambda^{-1})^\beta}_\gamma
d{\Lambda^\gamma}_\alpha.
\end{eqnarray}
Because of the non-invariant transformation law  of  ${\Gamma_\alpha}^\beta$
 as Eq. (\ref{g5}) shows it will not vanish in any other frame connected to the first by a
LLT.\vspace{0.5cm}\\
 \underline{ii}) The connection ${\Gamma_\alpha}^\beta$ plays an
essential task in the teleparallel framework. This task describes
the inertial causes which occur from the selection of the reference
system \cite{LOP}. The  evolvements of this inertial in many
situations lead to non-physical results of the total energy of the
system. Therefore, the function of the teleparallel connection, is to
deviate the inertial involvement  from the really gravitational
one. Because of the teleparallel curvature is vanishing identically, the connection is a
``pure gauge'', that is \begin{equation}\label{g6}
{\Gamma_\alpha}^\beta={(\Lambda^{-1})^\beta}_\gamma d
{\Lambda^\gamma}_\alpha.\end{equation} The  Weitzenb\"ock connection
forever has the form (\ref{g6}). The   translational momentum of Lagrangian (\ref{g1}) has the
form \cite{LOP}
\begin{eqnarray}\label{g7}
&& \widetilde{ {\cal H}}_\alpha=\frac{1}{16\pi}{\widetilde{\Gamma}}^{\beta
\gamma}\wedge  \eta_{\alpha \beta \gamma}, \nonumber\\
&&{\Gamma_\alpha}^\beta \stackrel {\rm def.}{=} {\widetilde
{\Gamma}_\alpha}^\beta -{{\cal K}_\alpha}^\beta,\end{eqnarray} where ${\widetilde
{\Gamma}_\alpha}^\beta $  is the purely Riemannian connection and
${\cal K}^{\mu \nu}$ is the contorsion 1-form that is connected to the
torsion by the relation
\begin{equation}\label{g8} {\cal T}^\alpha  \stackrel {\rm def.}{=}  {{\cal K}^\alpha}_\beta \wedge
\vartheta^\beta.\end{equation}
The teleparallel model (\ref{g1}) belongs to the class of quasi-invariant
theories. One can easily show that under a change of the coframe
$\vartheta'^{\beta} =\left({\Lambda(x)^\alpha}_\beta \right)
\vartheta'^{\beta}$, the Lagrangian  (\ref{g1}) changes by a total derivative:
\begin{equation}\label{g9}
\tilde{ {\cal V}}(\vartheta')=\tilde{ {\cal V}}(\vartheta)-\frac{1}{16\pi}d\left[{\left(\Lambda^{-1}\right)^\alpha}_\beta
d {\Lambda^\beta}_\gamma
 \wedge {\eta^\gamma}_\alpha \right]. \end{equation}
Besides to (\ref{g9}), it is an easy task to check  that  Eq. (\ref{g7}) varies
 like \begin{eqnarray}\label{g10}
 \tilde{{\cal H}'_\alpha}(\vartheta')&=& {\left(\Lambda^{-1}\right)^\beta}_\alpha
\tilde{ {\cal H}_\beta}(\vartheta)\nonumber\\
&-&\frac{1}{16\pi}d\left[{\left(\Lambda^{-1}\right)^\beta}_\alpha
{\left(\Lambda^{-1}\right)^\nu}_\gamma d {\Lambda^\gamma}_\mu \wedge
{{\eta_\beta}^\mu}_\nu \right],\nonumber\\
 \tilde{{\cal E}'_\alpha}(\vartheta')&=&
{\left(\Lambda^{-1}\right)^\beta}_\alpha
 \tilde{{\cal E}_\beta}(\vartheta)+d{\left(\Lambda^{-1}\right)^\beta}_\alpha \wedge
 \tilde{ {\cal H}_\beta}(\vartheta)\nonumber\\
 &-&\frac{1}{16\pi}d\left[{\left(\Lambda^{-1}\right)^\beta}_\alpha
 {\left(\Lambda^{-1}\right)^\nu}_\gamma d {\Lambda^\gamma}_\mu \wedge {{\eta_\beta}^\mu}_\nu \right].
\end{eqnarray}

The total conserved charge in the TEGR theory  has the form \cite{ORP}
\begin{equation}\label{g11} \tilde{{\cal
J}}(\xi,\vartheta)=\frac{1}{16\pi}\int_{\partial S}\xi^\alpha \tilde{ {\cal H}}_\alpha=\frac{1}{16\pi}\int_{\partial S}
\xi^\alpha \tilde{\Gamma}^{\beta \gamma} \wedge \eta_{\alpha \beta \gamma},
\end{equation}
with $\xi^\alpha=\xi \rfloor \vartheta^\alpha$.
Using  LLT Eq. (\ref{g11}) changes as
\begin{eqnarray}\label{g12}
\tilde{{\cal J'}}(\xi,\vartheta)&=&\tilde{{\cal J}}(\xi,\vartheta)-\frac{1}{16\pi}\int_{\partial S}
\xi^\alpha {\left(\Lambda^{-1}\right)^\nu}_\gamma d{\Lambda^\gamma}_\mu \wedge {{\eta_\alpha}^\mu}_\nu\nonumber\\
&=&\frac{1}{16\pi}\int_{\partial S}
\xi^\alpha \left(\tilde{\Gamma}^{\beta \gamma}-\bar{\Gamma}^{\beta \gamma}\right)\wedge \eta_{\alpha \mu \nu},
\end{eqnarray}
with ${\bar{\Gamma}_\mu}^\nu= \left(\Lambda^{-1}\right)^\nu_\gamma d{\Lambda^\gamma}_\mu$. The vector $\xi=\zeta^i \partial_i$  is independent  of the choice of the frame while its components, i.e., $\xi^\alpha$ change as a vector. Now let use apply Eq. (\ref{g11}) to the tetrad (\ref{tetrad}). Using the spherical local coordinates $(t,r,\theta, \varphi)$ the binary solution, by proceeding the following transformation $(z=r \cos\theta,~\rho=r \sin\theta)$. Recalling the vierbein (\ref{tetrad}), the frame is described by the coframe components:
\begin{eqnarray}\label{g13}
{\vartheta_1}^{\hat{0}}&=& e^{\frac{\nu(r,\theta)}{2}}
 dt,\\
{\vartheta_1}^{\hat{1}}&=&e^{\frac{\mu(r,\theta)}{2}}
dr,\\
{\vartheta_1}^{\hat{2}}&=&re^{\frac{\mu(r,\theta)}{2}}
d\theta,\\
{\vartheta_1}^{\hat{3}}&=&r\sin\theta e^{\frac{-\nu(r,\theta)}{2}} d\varphi.
\end{eqnarray}
If we take coframe  (\ref{g13}) we get the non-vanishing components   of the Riemannian connection
${\tilde{\Gamma}_\alpha}^\beta$  in the form
\begin{eqnarray}\label{g14}
{\tilde{\Gamma}_0}^1 &=&\frac{\nu_r}{2}e^{\frac{\nu(r,\theta)-\mu(r,\theta)}{2}}dt,\\
{\tilde{\Gamma}_0}^2 &=&\frac{\nu_\theta}{2}e^{\frac{\nu(r,\theta)-\mu(r,\theta)}{2r}}dt,\\
{\tilde{\Gamma}_1}^2 &=&\frac{\mu_\theta dr-2rd\theta--r^2\mu_r d\theta }{2r},\\
{\tilde{\Gamma}_1}^3 &=&-\sin\theta\frac{(2-r\nu_r )}{2}e^{\frac{-\nu(r,\theta)-\mu(r,\theta)}{2r}}d\varphi,\\
{\tilde{\Gamma}_2}^3 &=&-\sin\theta\frac{2\cot\theta-\nu_\theta }{2}e^{\frac{-\nu(r,\theta)-\mu(r,\theta)}{2r}}d\varphi.
\end{eqnarray}
Using (\ref{g14})   as
well as the Riemannian connection
${\tilde{\Gamma}_\alpha}^\beta$  and substitute into (\ref{g7}) we finally get
\begin{eqnarray}\label{g15}
\widetilde{ {\cal H}}_{\hat{0}} & = &\sin\theta e^{\frac{-\nu(r,\theta)}{2}}\Biggl\{ r [4+r(\mu_r-\nu_r)]d\theta\wedge d\varphi\nonumber\\
&-& [2\cot\theta-\nu_\theta +\mu_\theta] dr\wedge d\varphi\Biggr\}\,\nonumber\\
\widetilde{ {\cal H}}_{\hat{1}} & = &  2\cos\theta e^{\frac{-\mu(r,\theta)}{2}} dt\wedge d\varphi,\nonumber\\
\widetilde{ {\cal H}}_{\hat{2}} & = & 2\sin\theta e^{\frac{-\mu(r,\theta)}{2}} d\varphi\wedge dt,\nonumber\\
\widetilde{ {\cal H}}_{\hat{3}} & = & -e^{\frac{\nu(r,\theta)}{2}}\frac{r[2+r\nu_r +r\mu_r]d\theta\wedge dt+[\nu_\theta  +\mu_\theta]dt\wedge dr}{r}. \nonumber\\
\end{eqnarray}
For the BBHs solution we have for a vector field $\xi =\xi ^\alpha e_\alpha =\zeta ^i\partial _i$ with constant holonomic components, $\zeta ^i$, in the coordinate system used in (\ref{g13}), the direct evaluation of the integral (\ref{g11}) yields the following infinite total conserved charge
\begin{eqnarray}\label{g16}
{\cal J}& =&\frac{1}{16\pi}\int_{\partial S}\xi^\alpha \tilde{ {\cal H}}_\alpha
=\left(m_1+m_2-r\right)\zeta^0\nonumber\\
 &+& \left(\frac{m_2^2-m_1^2-2m_1 m_2}{3r}\right)\zeta^0\nonumber\\
 &+& \frac{4~d^2}{15}\left(\frac{m_2^2-m_1^2+2m_1 m_2}{r^3}\right)\zeta^0+O\left(\frac{1}{r^4}\right).
\end{eqnarray}
One possible solution of the above unfamiliar result is the use of regularized form of  total charges
\begin{equation}\label{g17}
\tilde{{\cal J}}(\xi,\vartheta)=\frac{1}{16\pi}\int_{\partial S}
 \xi^\alpha \left({\tilde{\Gamma}}^{\beta \gamma}-\tilde{\Gamma}^{\beta \gamma}_{[m_1=0,\ m_2=0]}\right) \wedge \eta_{\alpha \beta
 \gamma}.
\end{equation}
Using (\ref{g17}) we get the total conserved quantities in the form
\begin{eqnarray}\label{g18}
{\cal J} &=&\frac{1}{16\pi}\int_{\partial S}\xi^\alpha \tilde{ {\cal H}}_\alpha
=\left(m_1+m_2\right)\zeta^0\nonumber\\
&+&\left(\frac{m_2^2-m_1^2-2m_1m_2}{3r}\right)\zeta^0\nonumber\\
 &+& \frac{4~d^2}{15}\left(\frac{m_2^2-m_1^2+2m_1 m_2}{r^3}\right)\zeta^0+O\left(\frac{1}{r^4}\right),
\end{eqnarray}
which is a consistent with the results in \cite{Mj}. It is clear that the leading term is consistent with the case of Schwarzschild field, which presents the total mass of the system $m_1 + m_2$.
\subsection{Orbital decay}
The analysis of the total conserved quantities are combined only with the $\zeta^{0}$-component, while other terms vanish. Consequently, we conclude that the total conserved quantities are purely an energy. Since we have assumed that the separation between the two point masses is constant, we do not expect any energy loss in the orbit. But in order to estimate an initial guess for the energy loss by a binary system due to its gravitational radiation. We consider the case $m_{1}=m_{2}=m$ and the separation between the two point masses to be a function of time, i.e. $d\equiv d(t)$. Using (\ref{g18}) the energy contained by the system $E(t)$ can be written as
\begin{equation}\label{Energy}
    E(t)\approx 2m-\frac{2}{3}\frac{m^2}{r}+\frac{8}{15}\frac{m^2 d(t)^2}{r^3}.
\end{equation}
The rate of the energy loss by the binary system is
\begin{equation}\label{energyloss}
    \dot{E} \approx \frac{16}{15}\frac{m^2~d}{r^3}~\dot{d}.
\end{equation}
Using (\ref{Energy}) and (\ref{energyloss}) we evaluate
\begin{equation}\label{GW}
    \frac{\dot{E}}{E}\approx 2 \frac{\dot{d}}{d},
\end{equation}
which agrees with the Keplerian treatment of orbital period decay \cite{ST86}. It is worth to mention that the solution is an exact
one for which we can find such quantities of physical interest as radiation patterns without assuming a weak field initial data.
\section{Conclusion}\label{S7}
In the frame of the TEGR theory, we have derived an exact solution of BBHs space-time. The field equations enforce the Ricci scalar to vanish, while they prevent the vanishing of the torsion scalar. This property has been used to visualize the space-time of the BBHs field. This fact can serve in the torsion vs curvature singularity analysis. As a matter of fact, in the Riemannian picture, all the invariants due to the Ricci tensor vanish, while the only non-vanishing components of the Weyl (dual) tensor are the electric components. So we used the invariants $W1R$ and $W2R$ in this study. Additionally, we used the teleparallel scalar, in the Weitzenb\"{o}ck picture, to analyze the singularities. The results identify the physical singularities at $r_{\pm}=0$. However, in our case here, the torsion contours appear as repulsive lines, which might afford a physical explanation to the stress keeping the system balanced.

We have calculated the acceleration tensor to study the inertial acceleration required to keep the frame in a certain inertial state in the BBHs space-time so that in static frames the the inertial acceleration is precisely opposing the gravitational acceleration. We have identified repulsive gravity zones via the acceleration tensor in the inter-medium keeping the system balanced. In this way, we do not need to introduce a strut or membrane to interpret the configuration as in the GR theory.

We have studied the conserved quantities of the BBHs gravitational field. The regularized total charge calculations have been used to estimate the rate of the energy loss due to emission of gravitational radiation by the system.

In conclusion, we would like to point out that the non-vanishing torsion scalar of the vacuum solution of BBHs space-time can be used to study the singularities of this solution. Also, the modification of the TEGR Lagrangian by adding a scalar field Lagrangian. In some cases the scalar field can be induced directly from the torsion \cite{HN15}, which can be used to interpret the repulsive gravity zones near the naked singularities and in the inter-medium of the BBHs space-time. Actually, in this paper we just alarm the importance of studying these types of solutions in the TEGR framework. However, the singularity and the gravitational energy-momentum tensor analysis in the TEGR framework still needs more investigations.
\subsection*{Acknowledgments}
This article is partially supported by the Egyptian Ministry of Scientific Research under project No. 24-2-12.
\appendix
\section{Notation}\label{app:not}
As is known, the exterior product is indicated by $\wedge$, however the interior product of a vector $\xi$ and a p-form $\Psi$ is indicated by $\xi\rfloor \Psi$. The vector basis dual to the frame 1-forms
$\vartheta^{\alpha}$ is indicated by $e_\alpha$ and they fulfil $e_\alpha \rfloor \vartheta^{\beta}={\delta}_\alpha^\beta$. By using local coordinates $x^i$, we get $\vartheta^{\alpha}=h^\alpha_i
dx^i$ and $e_\alpha=h^i_\alpha \partial_i$ with $h^\alpha_i$ and $h^i_\alpha $ are the covariant and contravariant components of the tetrad field.  The volume  4-form is defined by $\eta \stackrel {\rm def.}{=} \vartheta^{\hat{0}}\wedge \vartheta^{\hat{1}}\wedge \vartheta^{\hat{2}}\wedge\vartheta^{\hat{3}}.$  Additionally, by using  the interior product we define
\[\eta_\alpha \stackrel {\rm def.}{=} e_\alpha \rfloor \eta = \ \frac{1}{3!} \
\epsilon_{\alpha \beta \gamma \delta} \ \vartheta^\beta \wedge \vartheta^\gamma \wedge \vartheta^\delta,\]
with $\epsilon_{\alpha \beta \gamma \delta}$ is completely antisymmetric tensor and  $\epsilon_{0123}=1$.
\begin{eqnarray}
\eta_{\alpha \beta} &:=& e_\beta \rfloor \eta_\alpha =
\frac{1}{2!}\epsilon_{\alpha \beta \gamma \delta} \
\vartheta^\gamma \wedge \vartheta^\delta,\\
\eta_{\alpha \beta \gamma} &:=& e_\gamma
\rfloor \eta_{\alpha \beta}= \frac{1}{1!} \epsilon_{\alpha \beta
\gamma \delta} \ \vartheta^\delta,
\end{eqnarray}
that are the bases for 3-, 2- and 1-forms respectively. Finally,
\[\eta_{\alpha \beta \mu \nu} \stackrel {\rm def.}{=} e_\nu \rfloor \eta_{\alpha \beta \mu}=
e_\nu \rfloor e_\mu \rfloor e_\beta \rfloor e_\alpha \rfloor \eta,\]
is the Levi-Civita tensor density. The $\eta$-forms fulfil the following useful identities:
\begin{eqnarray}\label{g19}
\vartheta^\beta \wedge
\eta_\alpha & :=  & \delta^\beta_\alpha
\eta,\nonumber \\
\vartheta^\beta \wedge \eta_{\mu \nu}  &:=& \delta^\beta_\nu \eta_\mu-\delta^\beta_\mu \eta_\nu,
\nonumber\\
\vartheta^\beta \wedge \eta_{\alpha \mu \nu}  &:= & \delta^\beta_\alpha \eta_{\mu \nu}+\delta^\beta_\mu
\eta_{\nu \alpha}+\delta^\beta_\nu \eta_{ \alpha \mu}, \nonumber\\
\vartheta^\beta \wedge \eta_{\alpha \gamma \mu \nu}  & :=  & \delta^\beta_\nu \eta_{\alpha \gamma
\mu}-\delta^\beta_\mu \eta_{\alpha \gamma \nu }+\delta^\beta_\gamma \eta_{ \alpha \mu \nu}-\delta^\beta_\alpha
\eta_{ \gamma \mu \nu}.\qquad \end{eqnarray}
The line element $ds^2 := g_{\alpha \beta}\vartheta^\alpha \bigotimes \vartheta^\beta$ is fulfil by the
space-time metric $g_{\alpha \beta}$.

One can consider teleparallel geometry as a gauge theory of translation \cite{HMMN,h7,hs7,ORP,Ni}. In such geometry the coframe $\vartheta^\alpha$  plays the role of the gauge translational potential of the gravitational field.  General relativity can be reconstructed as the teleparallel theory. From geometric viewpoint, teleparallel gravity can be regarded as a particular case  of the metric-affine gravity whose coframe 1-form $\vartheta^\alpha$ and  Lorentz connection  are due to distant parallelism constraint ${R_\alpha}^\beta=0$ \cite{Ni,OP,ORP}. In such
geometry the torsion 2-form
\begin{eqnarray}\label{g20}
{\cal T}^\alpha&=&D\vartheta^\alpha=d\vartheta^\alpha+{\Gamma_\beta}^\alpha\wedge
\vartheta^\beta\nonumber\\
&=&\frac{1}{2}{{\cal T}_{\mu \nu}}^\alpha \vartheta^\mu
\wedge \vartheta^\nu=\frac{1}{2}{{\cal T}_{i j}}^\alpha dx^i \wedge
dx^j,\end{eqnarray}
occurs as the gravitational gauge field strength, ${\Gamma_\alpha}^\beta$ is the Weitzenb\"ock connection
1-form, $d$ being the exterior derivative and finally $D$ is the exterior covariant derivative. The torsion ${\cal T}^\alpha$ can be divided into three irreducible parts: the tensor part, the trace, and the
axial trace, provided by \cite{h7,hs7,BV,Kt,KST,LOP}
\begin{equation}\label{g21}
{^ {\tiny{( 1)}}{\cal T}^\alpha} := {\cal T}^\alpha-{^ {\tiny{( 2)}}{\cal T}^\alpha}-{^ {\tiny{( 3)}}{\cal T}^\alpha},
\end{equation}
with
\begin{equation}
{^  {\tiny{( 2)}}{\cal T}^\alpha}:=\frac{1}{3} \vartheta^\alpha\wedge {\cal T},
\end{equation}
where ${\cal T}=\left(e_\beta \rfloor {\cal T}^\beta\right)$ and $e_\alpha \rfloor
{\cal T}={{\cal T}_{\mu \alpha}}^\mu$ vectors of torsion trace, and
\begin{equation}
{^  {\tiny{( 3)}}{\cal T}^\alpha} :=\frac{1}{3} e^\alpha\rfloor {\cal P},
\end{equation}
with ${\cal P}=\left(\vartheta^\beta \wedge {\cal T}_\beta\right)$ and $e_\alpha\rfloor P={\cal T}^{\mu \nu \lambda}\eta_{\mu \nu \lambda \alpha}$ the axial torsion trace.
\bibliographystyle{apsrev}
\bibliography{1507.07377}
\end{document}